\documentclass[pdflatex,sn-basic,Numbered]{sn-jnl}

\graphicspath{ {./paper-figures} }
\usepackage{graphicx}%
\usepackage{multirow}%
\usepackage{amsmath,amssymb,amsfonts}%
\usepackage{amsthm}%
\usepackage{mathrsfs}%
\usepackage[title]{appendix}%
\usepackage[normalem]{ulem}
\usepackage{xcolor}%
\usepackage{textcomp}%
\usepackage{manyfoot}%
\usepackage{booktabs}%
\usepackage{physics}
\usepackage{algorithm}%
\usepackage{algorithmicx}%
\usepackage{algpseudocode}%
\usepackage{listings}%
\usepackage{caption, subcaption}

\newcommand\hl[1]{\textcolor{black}{#1}}

\theoremstyle{thmstyleone}%
\theoremstyle{thmstyletwo}%

\theoremstyle{thmstylethree}%

\begin{document}

\title[Article Title]{Using Diffusion Models to Generate Synthetic Labeled Data for Medical Image Segmentation}


\author[1]{\fnm{Daniel} \sur{Saragih}}\email{daniel.saragih@mail.utoronto.ca}

\author[1,2]{\fnm{Atsuhiro} \sur{Hibi} \dgr{MSc}}\email{atsuhiro.hibi@mail.utoronto.ca}

\author*[1,2,3]{\fnm{Pascal N.} \sur{Tyrrell} \dgr{PhD}}\email{pascal.tyrrell@utoronto.ca}

\affil[1]{\orgdiv{Department of Medical Imaging}, \orgname{University of Toronto}, \orgaddress{\street{263 McCaul Street}, \city{Toronto}, \postcode{M5T 1W7}, \state{ON}, \country{Canada}}}

\affil[2]{\orgdiv{Institute of Medical Science}, \orgname{University of Toronto}, \orgaddress{\city{Toronto}, \state{ON}, \country{Canada}}}

\affil[3]{\orgdiv{Department of Statistical Sciences}, \orgname{University of Toronto}, \orgaddress{\city{Toronto}, \state{ON}, \country{Canada}}}

\abstract{
Medical image analysis has become a prominent area where machine learning has been applied. However, high quality, publicly available data is limited either due to patient privacy laws or the time and cost required for experts to annotate images. In this retrospective study, we designed and evaluated a pipeline to generate synthetic labeled polyp images for augmenting medical image segmentation models with the aim of reducing this data scarcity.  In particular, we trained diffusion models on the HyperKvasir dataset, comprising 1000 images of polyps in the human GI tract from 2008 to 2016. Qualitative expert review, Fr\'echet Inception Distance (FID), and Multi-Scale Structural Similarity (MS-SSIM) were tested for evaluation. Additionally, various segmentation models were trained with the generated data and evaluated using Dice score and Intersection over Union. We found that our pipeline produced images more akin to real polyp images based on FID scores, and segmentation performance also showed improvements over GAN methods when trained entirely, or partially, with synthetic data, despite requiring less compute for training. Moreover, the improvement persists when tested on different datasets, showcasing the transferability of the generated images. 
}




\maketitle

\section*{Acknowledgments}
 We would like to thank \textbf{Ali Geramy} for his assistance in setting up the computing environment for training. And, we also would like to thank \textbf{Felipe Castillo}, for their time and effort in reviewing the synthetic images as part of the expert review.

\section{Introduction}\label{sec:Introduction}

In recent decades, artificial intelligence (AI) has been growing in prominence, and its numerous uses have begun to be realized, including those in healthcare \cite{jiangArtificialIntelligenceHealthcare2017,yuArtificialIntelligenceHealthcare2018}. For example, AI has been used to simulate molecular dynamics, discover drugs, and diagnose diseases \cite{yuArtificialIntelligenceHealthcare2018}. Among these applications, medical image analysis has become a prominent area where AI has been applied \citep{willeminkPreparingMedicalImaging2020,thambawitaSinGANSegSyntheticTraining2022}. 

The ML models used in these applications learn from data. Of particular interest in this paper are segmentation models and the data used to train them. Segmentation models classify pixels in the image into regions, and in the case of medical imaging, this task entails delineating malignancies, functional tissues, or organs of interest \citep{pratondoIntegratingMachineLearning2017,xuMedicalBreastUltrasound2019,navarroShapeAwareComplementaryTaskLearning2019}. The quality and quantity of the data is a significant factor in the model’s performance. However, publicly available data is limited either due to patient privacy laws or the time and cost required for experts to annotate the images \citep{willeminkPreparingMedicalImaging2020}. 

The addition of synthetically generated labeled data is a possible approach to tackle the lack of training data as it would overcome the need for annotations by experts. Indeed, GANs have been used to generate training data and labels in the context of medical images to train detection and segmentation models \citep{shinAbnormalColonPolyp2018,lindnerUsingSyntheticTraining2019}. Moreover, the advent of Denoising Diffusion Probabilistic Models (DDPM) or diffusion models \cite{hoDenoisingDiffusionProbabilistic2020} present a new method of generating novel images which has shown remarkable results in creating photorealistic images \cite{borjiGeneratedFacesWild2023}. 

The purpose of this paper was to explore the capabilities of diffusion models in generating synthetic data for medical image segmentation. We looked at some recent work to motivate our objectives. Diffusion models have been used to successfully augment image datasets in \cite{khaderDenoisingDiffusionProbabilistic2023,trabuccoEffectiveDataAugmentation2023c} for medical imaging and a general classification task. In particular, \citet{trabuccoEffectiveDataAugmentation2023c} showed an improvement in a few-shot classification task. Moreover, \citet{thambawitaSinGANSegSyntheticTraining2022} used styling to improve segmentation performance on the HyperKvasir dataset \cite{borgliHyperKvasirComprehensiveMulticlass2020}.
The contributions of the paper are as follows:
\begin{enumerate}
    \item We explored the use of clustering as a pre-processing step prior to training the diffusion model. This was done to overcome the large variation between images in the dataset which the diffusion model struggles to represent.
    \item Our diffusion models had a fourth channel input which was used to generate the segmentation mask. This was done to ensure that the model generated images with corresponding masks simultaneously, alleviating the need for image annotation by experts or a separate model.
    \item We used the RePaint technique introduced by \citet{lugmayrRePaintInpaintingUsing2022} on the polyps as well as styling \cite{kwonDiffusionbasedImageTranslation2023,yangZeroShotContrastiveLoss2023a} to generate realistic synthetic images that possesses sufficient variation to be used for training.
\end{enumerate}

\noindent
The code used in this paper is available on GitHub.\footnote{\url{https://github.com/dsaragih/diffuse-gen}}

\section{Methods}\label{sec:Methods}

\begin{figure}
  \centering
  \includegraphics[width=0.9\textwidth]{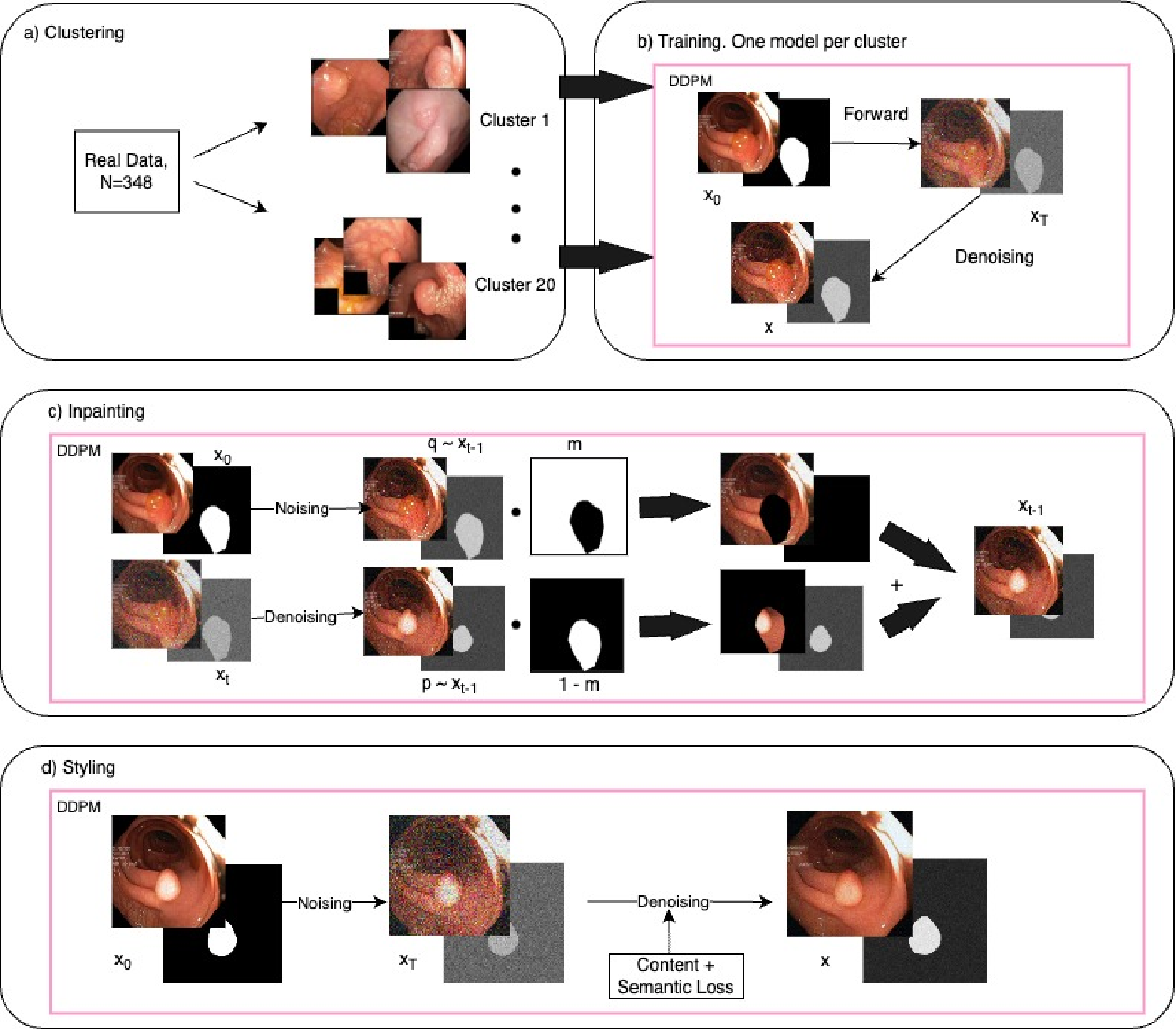}
  \caption{Schematic of our method. Given a dataset of real images and masks (here \(N = 348\)), we \textbf{a)} cluster the images and masks by similarity into \(K = 20\) clusters, \textbf{b)} train a diffusion model on each cluster, \textbf{c)} use the trained diffusion model and the inpainting technique to generate synthetic images and masks, and \textbf{d)} style the synthetic images by guiding the reverse diffusion process with various losses.}
  \label{fig:schematic}
\end{figure}

Here we present the methods of our study where we sought to design and evaluate a pipeline for generating synthetic labeled polyp images for augmenting automatic segmentation models. An illustration of the process is shown in Figure \ref{fig:schematic}.
 
\subsection{Dataset}
The dataset we used is the publicly available HyperKvasir \cite{borgliHyperKvasirComprehensiveMulticlass2020} dataset, which we retrieved at \url{https://osf.io/mh9sj/}, from which we obtained 348 polyp images with a corresponding segmentation mask annotated by medical experts.  This dataset, which consists of polyp images in the human GI tract, was collected from 2008 to 2016 during clinical endoscopies and will be the dataset used to train our pipeline \cite{borgliHyperKvasirComprehensiveMulticlass2020}. Specifically, our goal was to use this dataset as a case study due to the time and resources required to train our pipeline. However, it's expected that our approach may be generalized to other segmentation datasets.

\begin{figure}
    \centering
    \includegraphics[width=\textwidth]{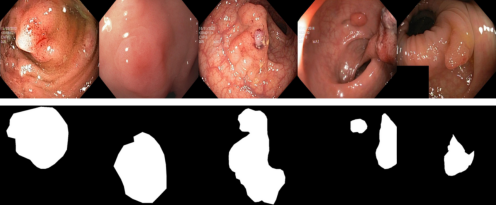}
    \caption{A few samples of the HyperKvasir dataset.}
    \label{fig:hyperkvasir}
\end{figure}

A few samples of the dataset are shown in Figure \ref{fig:hyperkvasir}. The polyp images are RGB images, whereas the masks are single-channeled images. The mask colours the region with polyp white, while the surrounding regions are coloured black. This dataset was used to train our generative diffusion models, and compare the performance of the segmentation models depending on the type of training data used.

\hl{In addition, we used the CVC-ClinicDB dataset \cite{bernalWMDOVAMapsAccurate2015} and the PolypGen dataset \cite{aliMulticentrePolypDetection2023} to evaluate the performance of the segmentation models. Specifically, they were used to test the generalization of the models trained on the HyperKvasir dataset. The CVC-ClinicDB dataset contains 612 images with corresponding masks, while PolypGen includes both single frame and sequence data with 3762 annotated polyp label; we restrict ourselves to the 1537 single frame images.}\footnote[1]{https://github.com/DebeshJha/PolypGen}\,\footnote[2]{https://www.kaggle.com/datasets/balraj98/cvcclinicdb}

\subsection{Image Preprocessing}

\begin{figure}
  \centering
  \includegraphics[width=0.8\textwidth]{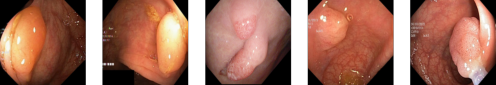}
  \includegraphics[width=0.8\textwidth]{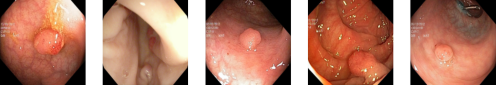}
  \includegraphics[width=0.8\textwidth]{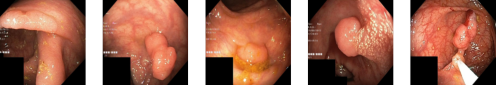}
  \caption{Example of clustering results. Each row consists of images from a cluster.}
  \label{fig:cluster}
\end{figure}

\hl{We first used a variant\footnote[3]{https://github.com/mseitzer/pytorch-fid} of the PyTorch Inception-v3 \cite{szegedyRethinkingInceptionArchitecture2016} model pretrained on ImageNet to obtain a feature set for all image-mask pairs in the dataset. To do so, we use the most straightforward method: we pass each image into the model and later the mask separately. Afterwards, we concatenate the image features and mask features into a single vector.} On this newly-formed vector, we used scikit-learn's KMeans algorithm \hl{to obtain \(K \in \{10, 20, 30, 40\}\) clusters of images. Throughout this process, we use a NVIDIA RTX A6000 GPU, requiring no more than 5 minutes to form the clusters}. See Figure \ref{fig:cluster} for examples of images in such clusters; each row depicts images in one cluster. The images and masks were then resized to \(256 \times 256\) and pixel values scaled to \([-1, 1]\). 

\subsection{Training}

We trained our diffusion model one by one for each cluster we created. Although clustering adds to the number of models trained and hence requires more compute, in our use case of small datasets, the number of clusters needed to sensibly group similar images is small. Hence, the extra compute and time is mild in practice.

In order to train the model to generate labels, we concatenated the mask as a 4th input channel. Our method inherited from the diffusion model implementation of \cite{kwonDiffusionbasedImageTranslation2023} and the generation technique of \cite{thambawitaSinGANSegSyntheticTraining2022}.

The model was trained for \hl{\(T = 110,000\) iterations, meaning that a total of \(T\) training images were passed through the model,} with hyperparameters given in Supplementary Information B.5. Moreover, we trained one additional model on the entire dataset to ablate the effectiveness of clustering. This model was trained on the same parameters for \(T = 200,000\) iterations. The models were trained on Mist, a SciNet GPU cluster \cite{lokenSciNetLessonsLearned2010,ponceDeployingTop100Supercomputer2019}; each node of the cluster has 4 NVIDIA V100-SMX2-32GB GPU. The training time for each model was at most 16 hours, with variation resulting from cluster size.

\subsection{Image Generation}

After training the generative diffusion models, we generated 3 random samples for each real image using the inpainting technique introduced in \cite{lugmayrRePaintInpaintingUsing2022} (see Supplementary Information A.2 for details) with the parameters given in Supplementary Information B.5. This procedure used the polyp mask to obscure the region of the image coloured white in the mask. The diffusion model is thus tasked with filling in this gap. The upside of this procedure as compared to generating the full image is that we conditioned the diffusion model on the surrounding region. This significantly contributed to the realism of the image, while also injecting sufficient variation to the image.

\hl{More specifically, suppose we would like to inpaint images in cluster \(i\). We first randomly select \(j \sim \mathcal{U}\{1, \ldots, K\}\) to choose a cluster model. Using the terminology of \citet{lugmayrRePaintInpaintingUsing2022}, the images of cluster \(i\) are the source images. Since the model was trained on cluster \(j\), the inpainted polyp is in the style of cluster \(j\). Without this randomness, models trained on smaller clusters would have very limited variation, often resulting in their original images unchanged. By introducing this randomness, we ensure greater variation and a balance between smaller and larger clusters.}

Note that a number of the polyp masks failed to obscure the entire polyp, and thus, the model recognized the polyp in its training set and returned the original image. To mitigate this, we dilated the polyp mask by 20 pixels which reduced the number of such cases.

Finally, we used a combination of the styling methods in \cite{kwonDiffusionbasedImageTranslation2023,yangZeroShotContrastiveLoss2023a} to generate the final synthetic images. In particular, after sampling \(348 \times 3\) inpainted images, the style transfer procedure \cite{kwonDiffusionbasedImageTranslation2023,yangZeroShotContrastiveLoss2023a} was applied to each image. A NVIDIA RTX A6000 was used for this procedure, which was carried out by the previously trained diffusion model. Each image required about 1 minute to style. The hyperparameters of this procedure are given in Supplementary Information B.5.

We combined the two methods by applying \cite{yangZeroShotContrastiveLoss2023a} for the content loss and \cite{kwonDiffusionbasedImageTranslation2023} for the style loss. This takes advantage of the improvements in \cite{yangZeroShotContrastiveLoss2023a} while using a style loss that does not require a text prompt. The total loss is the sum of the content and style loss; refer to Supplementary Information A.3 for loss details.

\subsection{Intrinsic Evaluation}
To evaluate our samples, we did a qualitative expert review, and computed the Fr\'echet Inception Distance (FID) \cite{heuselGANsTrainedTwo2017} and Multi-Scale Structural Similarity (MS-SSIM) \cite{wangMultiscaleStructuralSimilarity2003} between the output images and the real data.

The FID \cite{heuselGANsTrainedTwo2017} and MS-SSIM \cite{wangMultiscaleStructuralSimilarity2003} scores measure the distance between the output images and the dataset, as well as the variance of the output images. \citet{parmarAliasedResizingSurprising2022} showed that differences in image compression and resizing, among other factors, may induce significant variation in the FID scores. Hence, we used the method introduced in \cite{parmarAliasedResizingSurprising2022} to compute the FID.

For later convenience, we introduce the following convention when referring to each data type. Note that when the prefix or suffix is unnecessary, we omit it.
\begin{itemize}
  \item \textbf{Real}: Real images from the training set.
  \item \hl{\textbf{GAN-N}}: Synthetic images generated by SinGAN-Seg \cite{thambawitaSinGANSegSyntheticTraining2022}, downloaded at \url{https://osf.io/xrgz8/}. \hl{The suffix \(N\) denotes the number of images sampled, as per Section 2.4, \(N = 1, 2,\) or \(3\). Though some experiments were similarly conducted in the original paper, we re-ran them to ensure consistency.}
  \item \hl{\textbf{K-Diff-N}}: Images generated using the inpainting technique by a diffusion model trained on a single cluster. No styling applied. \hl{The prefix \(K\) denotes the cluster number, chosen from \(\{10, 20, 30, 40\}\).}
  \item \hl{\textbf{K-Styled-Diff-N}}: Same as \textbf{Diff} with styling applied.
  \item \hl{\textbf{Full-Diff-N}}: Images generated using the inpainting technique by a diffusion model trained on the full dataset. No styling applied.
  \item \hl{\textbf{Full-Styled-Diff-N}}: Same as \textbf{Full-Diff} with styling applied.
\end{itemize}

\begin{figure}
  \centering
  \includegraphics[width=0.4\textwidth]{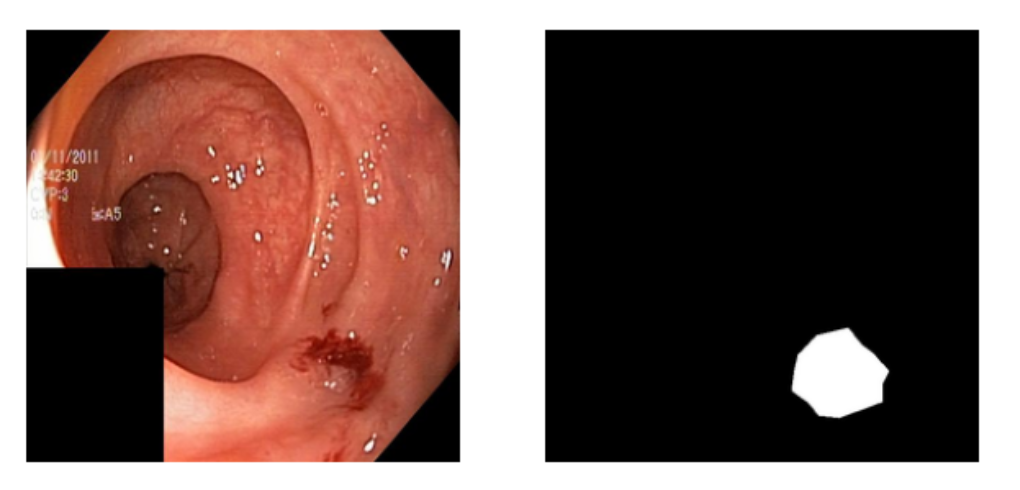}
  \hspace{35pt}
  \includegraphics[width=0.4\textwidth]{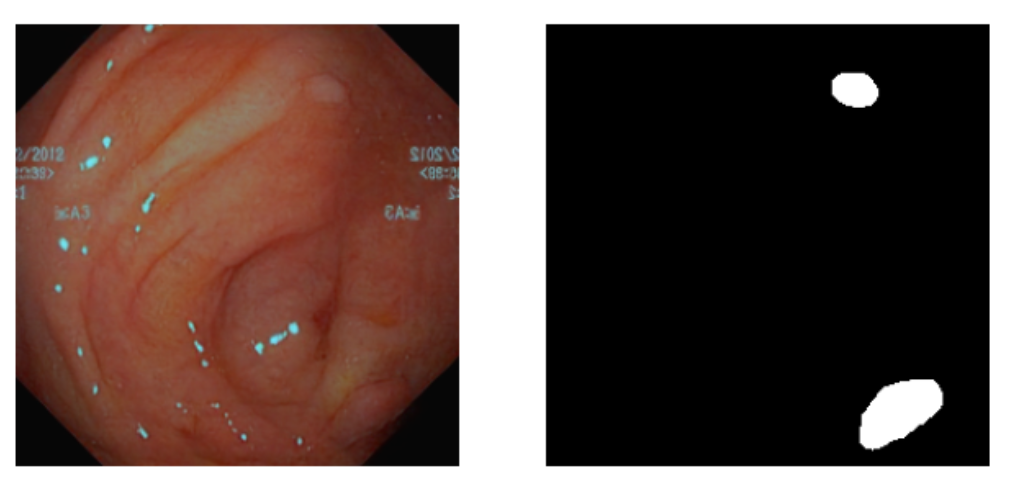}
  \caption{Example of survey images. Left pair is Real, whereas right pair is Fake.}
  \label{fig:survey}
\end{figure}

For our expert review of the generated image, we had a radiologist (with 5 years of experience) review 30 shuffled images and their masks side by side; see Figure \ref{fig:survey} for a sample. Out of these 30 images, 10 were real images, 10 were from \textbf{20-Diff-1}, and 10 were from \textbf{20-Styled-Diff-1}. See Supplementary Information B.1 for further details and results.

\subsection{\hl{Segmentation Experiments and Transfer Learning}}

Subsequently, we trained a polyp segmentation model \cite{thambawitaDivergentNetsMedicalImage2021} with the original and synthetic dataset to compare their mean IoU and DS with the target masks in order to evaluate their effect on model performance.
Specifically, we compared the performance of the model when trained under standard augmentation methods with synthetic data generated by \cite{thambawitaSinGANSegSyntheticTraining2022}, and with our synthetic data. We ran two sets of experiments: 
\begin{itemize}
  \item Full Synthetic Training: The model was trained on the entire synthetic dataset.
  \item Augmented Synthetic Training of Small Datasets: The model was trained on a dataset consisting of a subset (\(N = 16, 32, 64, 128\)) of real images augmented with synthetic images.
\end{itemize}

\subsubsection{Full Synthetic Training}
\hl{In the first experiment,} we used the model proposed by \citet{thambawitaDivergentNetsMedicalImage2021}; see Supplementary Information B.3 for details. We performed three runs, each corresponding to the three samples. \hl{The experiment \textbf{20-Diff-N}}, for example, means that \(N \times 348\) images were used in the training set, all chosen from the diffusion model output without styling. These were split into training (80\%) and validation (20\%) sets. 

For the test set, we used the HyperKvasir \cite{borgliHyperKvasirComprehensiveMulticlass2020} dataset at \url{https://datasets.simula.no/hyper-kvasir/}, which contains 1000 images with corresponding masks. We selected images not used in the 348-size training set, and randomly chose 200 of these images as the test set. \hl{Additionally, for our transfer learning evaluations, we used 200 randomly chosen images from CVC-ClinicDB \cite{bernalWMDOVAMapsAccurate2015} and PolypGen \cite{aliMulticentrePolypDetection2023}.} Tests were conducted on the best checkpoint, as determined by the validation IoU score.

\subsubsection{Augmented Synthetic Training of Small Datasets}
\hl{In the second experiment,} we experimented with the effect of augmentation on model performance when data is scarce. We followed the approach of \citet{trabuccoEffectiveDataAugmentation2023c} when performing augmentation. In particular, we fix \(\alpha \in (0, 1)\), and sample indices \(i, j\) uniformly \[
  i \sim \mathcal{U}({1, \ldots, N}), \quad j \sim \mathcal{U}({1, \ldots, M})
\]
where \(N\) is the number of images in the training set and \(M\) is the number of synthetic images per real image -- in our case \(N \in \{16, 32, 64, 128\}, M = 3\). In \citet{trabuccoEffectiveDataAugmentation2023c}, \(\alpha = 0.5\) was set as the probability a synthetic image \(\tilde{X}_{ij}\) generated from real image \(X_i\) would be added to the training set. However, as it is typically the case that we have more synthetic data per real image, instead of simple addition with probabilty \(\alpha = 0.5\), we add \(r=1, 2,\) or \(3\) synthetic images generated from \(X_i\). Because of resource and time limitations, we restrict our attention to \textbf{Styled-Diff} and \textbf{GAN}. The experiments used the same parameters as those in the Full Synthetic Training experiments.

\section{Results}\label{sec:Results}
\hl{Here, we present the experiemental results of our study. We restrict ourselves to presenting the results for cluster count 20 (hence, we omit the cluster prefix), with the exception of Section 3.3.2. Though the results may not be the best among all cluster counts, they are representative of the overall performance of the cluster model, and sufficient to demonstrate the highlighted trends. The results for other cluster counts are available in the supplementary material (see Supplementary Information B.2, B.4).}

\subsection{Image Generation}

\begin{figure}
  \centering
  \begin{subfigure}[b] {0.45\textwidth}
    \caption{}
    \includegraphics[width=\textwidth]{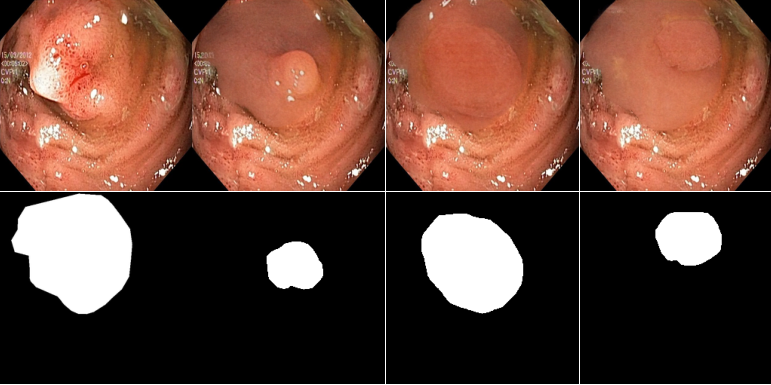}
    \label{fig:inpainting0}
  \end{subfigure}
  \begin{subfigure}[b] {0.45\textwidth}
    \caption{}
    \includegraphics[width=\textwidth]{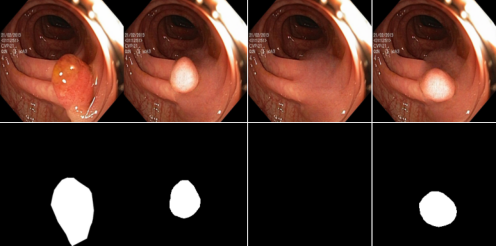}
  \label{fig:inpainting1}
  \end{subfigure}
  \begin{subfigure}[b] {0.45\textwidth}
    \caption{}
    \includegraphics[width=\textwidth]{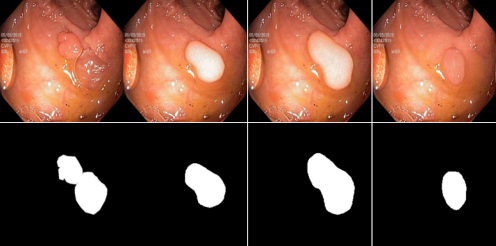}
  \label{fig:inpainting2}
  \end{subfigure}
  \begin{subfigure}[b] {0.45\textwidth}
    \caption{}
    \includegraphics[width=\textwidth]{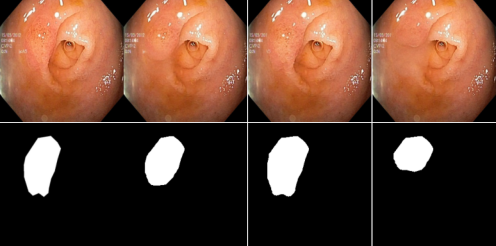}
  \label{fig:inpainting3}
  \end{subfigure}
  \caption{Example of inpainting results. Each subpart corresponds to a different source image; the first column is the source image, and the subsequent columns are synthetically generated using image inpainting, \bf{Full-Diff}.}
  \label{fig:inpainting}
\end{figure}

The real data points were used as input to the diffusion models, and the segmentation mask is used as the inpainting mask. Recall that we are using a four-channel model, the last used by the segmentation mask. Consequently, our inpainting procedure generated a new image and its corresponding segmentation, alleviating much work for image annotators. Four real training images and three corresponding inpainted images are depicted in Figure \ref{fig:inpainting}. The first column of each subfigure represents the original image; the mask of each image is directly beneath it.

\begin{figure}
  \centering
  \begin{subfigure}[b]{0.45\textwidth}
    \caption{}
    \includegraphics[width=\textwidth]{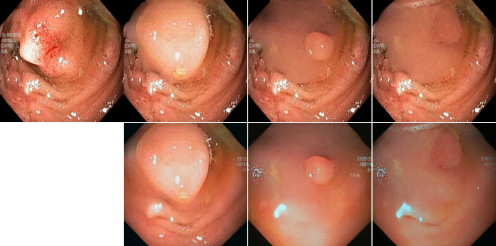}
    \label{fig:styled0}
  \end{subfigure}
  \begin{subfigure}[b]{0.45\textwidth}
    \caption{}
    \includegraphics[width=\textwidth]{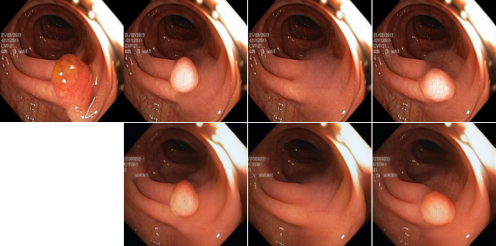}
    \label{fig:styled1}
  \end{subfigure}
  \begin{subfigure}[b]{0.45\textwidth}
    \caption{}
    \includegraphics[width=\textwidth]{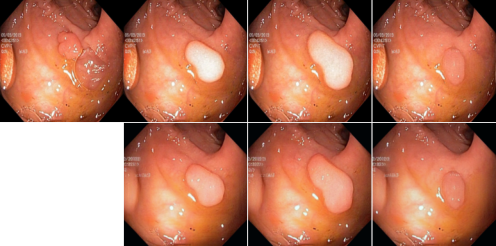}
  \label{fig:styled2}
  \end{subfigure}
  \begin{subfigure}[b]{0.45\textwidth}
    \caption{}
    \includegraphics[width=\textwidth]{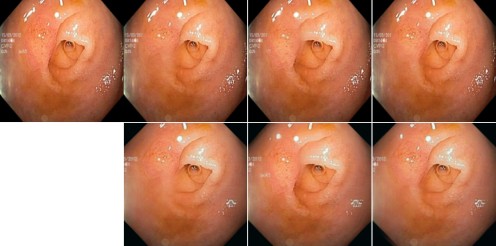}
  \label{fig:styled3}
  \end{subfigure}
  \caption{Example of styled results. Each subpart corresponds to a different source image; the first column is the source image, and the subsequent columns compare the image with (second row) and without styling (first row). The corresponding segmentation mask remains the same after this process.}
  \label{fig:styled}
\end{figure}

The styling results are depicted in Figure \ref{fig:styled}. The first row of each subfigure corresponds to the first row of each subfigure in Figure \ref{fig:inpainting}, whereas the second row is the styled image. By visual comparison in Figure \ref{fig:styled}, we see that styling improved the appearance of the synthetic images. This was done by smoothing out the inpainting region so as to remove the sudden changes in colour at the boundary of the inpainting region.

\subsection{Intrinsic evaluation}

The FID and MS-SSIM values are reported in Table \ref{table:fidnmsssim}. The FID of \textbf{Styled-Diff} (\(65.99 \pm 0.37\)) is significantly different from the other datasets. In particular, this means it is significantly closer to the real images than \textbf{GAN}, but significantly farther than the other variations on \textbf{Diff}. Moreover, the variations on \textbf{Diff} obtained a significantly lower FID than \textbf{GAN}, indicating that they are more alike the real image. It is also true for MS-SSIM that \textbf{Styled-Diff} (\(0.2346 \pm 0.0062\)) is significantly greater than the other datasets, including \textbf{Real}. Moreover, the MS-SSIM values of the synthetic data fall within \(0.05\) of the \textbf{Real} data, indicating that the variance of the synthetic data is similar to the real data. This is expected as the diffusion model is trained on the real dataset, and the inpainting technique may only add elements from other images in the dataset.

\begin{table}
  \centering
  \caption{Fr\'echet Inception Distance (FID) and Multi-Scale Structural Similarity (MS-SSIM) comparison between real and synthetic images. Images are shuffled, and each data type (including \textit{Real}) is compared with \textit{Real}. The computations were performed thrice.}
  \label{table:fidnmsssim}
  \begin{tabular}{cccccccccc}
  \toprule
  Data type & & &\multicolumn{3}{c}{FID} & \multicolumn{3}{c}{MS-SSIM} \\
  \cmidrule(r){4-6}
  \cmidrule(r){7-9}
   & & & Set 1 & Set 2 & Set 3 & Set 1 & Set 2 & Set 3 \\
  \midrule
  Real & & & & $-4.412 \cdot 10^{-5}$ & & & 0.1813 & \\
  GAN & & & 118.73 & 119.45 & 116.92 & 0.1986 & 0.2004 & 0.2037 \\
  Diff & & & 36.01 & 37.90 & 37.78  & 0.2081 & 0.2126 & 0.2096 \\
  Styled-Diff & & & 66.17 & 65.48 & 66.32 & 0.2304 & 0.2300 & 0.2433 \\
  Full-Diff & & & 40.59 & 40.77 & 40.27 & 0.2030 & 0.2058 & 0.2078 \\
  Full-Styled-Diff & & & 60.15 & 59.98 & 59.95 & 0.1971 & 0.1924 & 0.1969 \\
  \bottomrule
  \end{tabular}
\end{table}

We should be careful in interpreting the FID metric because our inpainting technique necessarily leaves parts of the original image intact. Our caution is supported by the fact that our pipeline without styling has a lower FID than the styled pipeline. For instance, in Figure \ref{fig:styled0}, the styled image shows a drastic change in the surroundings so as to better match the inpainted region. On one hand, this smoothening effect created a more natural appearance, but on the other, it increased the distance between the synthetic and real images. However, as we see in Figure \ref{fig:styled}, the images without styling contains noticeable sudden changes in colour at the inpainting region, resulting in an unnatural appearance. Thus, despite the visual improvement FID is higher after styling. Furthermore, for our purpose of training segmentation models, some deviation from the real images is desirable as we want to avoid overfitting to the real images.

\subsection{\hl{Segmentation experiments and Transfer Learning}}

\begin{table}
  \centering
  \caption{Performance evaluation of segmentation models entirely trained on synthetic data. \hl{The test datasets are HyperKvasir images not used in the generation, or images sampled from CVC-ClinicDB and PolypGen.}}
  \label{table:segmentation}
  \begin{tabular}{ccccccccc}
    \toprule
     & & & \multicolumn{2}{c}{HyperKvasir} & \multicolumn{2}{c}{\hl{CVC-ClinicDB}} & \multicolumn{2}{c}{\hl{PolypGen}} \\
    \cmidrule(r){4-5} \cmidrule(r){6-7} \cmidrule(r){8-9}
    Data type & & & Dice Score (\(\uparrow\)) & IoU (\(\uparrow\)) & \hl{Dice Score} & \hl{IoU} & \hl{Dice Score} & \hl{IoU}\\
    \midrule  
    Real & & & 0.8680 & 0.8085 & 0.7341 & 0.6604 & 0.6185 & 0.6284\\
    \midrule
    GAN-1 & & & 0.6553 & 0.5790 & 0.4724 & 0.4208 & 0.4699 & 0.4875 \\
    Diff-1 & & & 0.6566 & 0.5948 & 0.5845 & 0.5279 & 0.4351 & 0.4923 \\
    Styled-Diff-1 & & & \textbf{0.7444} & \textbf{0.6840} & \textbf{0.6363} & \textbf{0.6027} & \textbf{0.5669} & \textbf{0.5871}\\
    Full-Diff-1 & & & 0.6811 & 0.6149 & 0.5947 & 0.5400 & 0.4762 & 0.5112 \\
    Full-Styled-Diff-1 & & & 0.6514 & 0.5751 & 0.5533 & 0.5096 & 0.4543 & 0.4865 \\
    \midrule
    GAN-2 & & & 0.7062 & 0.6239 & 0.5425 & 0.5089 & 0.5005 & 0.5188 \\
    Diff-2 & & & 0.6626 & 0.5975 & 0.6315 & 0.5827 & 0.4934 & 0.5203 \\
    Styled-Diff-2 & & & \textbf{0.7728} & \textbf{0.7027} & \textbf{0.6414} & \textbf{0.5890} & \textbf{0.5444} & \textbf{0.5752}\\
    Full-Diff-2: & & & 0.6597 & 0.5983 & 0.5880 & 0.5011 & 0.4894 & 0.5016 \\
    Full-Styled-Diff-2 & & & 0.6077 & 0.5534 & 0.5721 & 0.5129 & 0.4923 & 0.4819 \\
    \midrule
    GAN-3 & & & 0.6960 & 0.6256 & 0.4945 & 0.4425 & 0.4911 & 0.5021 \\
    Diff-3 & & & 0.6516 & 0.5883 & 0.6335 & 0.5776 & 0.4694 & 0.4703\\
    Styled-Diff-3 & & & \textbf{0.8042} & \textbf{0.7396} & \textbf{0.6749} & \textbf{0.6124} & \textbf{0.5753} & \textbf{0.5728} \\
    Full-Diff-3 & & & 0.6085 & 0.5501 & 0.5904 & 0.5210 & 0.5066 & 0.5002 \\
    Full-Styled-Diff-3 & & & 0.5649 & 0.5213 & 0.5117 & 0.4623 & 0.4824 & 0.4698 \\
    \bottomrule
  \end{tabular}
\end{table}

\subsubsection{Full Synthetic Training}
\hl{Our first experiment concerns full synthetic training, shown in Table \ref{table:segmentation}. We trained the model only using generated data and tested using real data not used in the generation from HyperKvasir, as well as images from CVC-ClinicDB and PolypGen. We see that  across \(N = 1, 2, 3\) and all datasets, \textbf{Styled-Diff} has the highest DS and IoU scores, increasing with \(N\). It trails behind the \textbf{Real} metrics, with the difference narrowing on the transfer datasets. \textbf{Styled-Diff} performed better than \textbf{GAN} data in all cases, indicating that the generated data is more representative of true polyp images. Moreover, styling resulted in consistent improvements in the DS and IoU scores. Indeed, excepting \(N = 2, 3\) for CVC-ClinicDB, the styled counterparts obtained at least a \(0.05\) gain in both metrics. The effect of clustering can also be observed: \textbf{Styled-Diff} often outperforms \textbf{Full-Styled-Diff} by \(0.1\) in both metrics.}

\subsubsection{Augmentation of Small Datasets} 
\hl{Our second experiment concerns augmenting small datasets, given in Table \ref{table:small_augmentation}. When augmenting small datasets, the synthetic data performed better than Real with standard augmentation techniques (Supplementary Information B.3). This is because the synthetic data incorporated elements from images not in the training set, and thus contributed new information. Indeed, this effect is most noticeable when the training set is starved for data, such as when \(N = 16\), where the largest improvement in the DS and IoU scores occurs for both \textbf{Styled-Diff} and \textbf{GAN}. The effect diminishes as the training set size increases, however, \textbf{Styled-Diff} still ekes out a better score on the transfer datasets when \(N = 128\). It is also worth noting that \textbf{Styled-Diff} consistently outperforms \textbf{GAN} in IoU scores, but, at times, has a slightly lower DS score. Since IoU more heavily penalizes misclassification, this indicates that the \textbf{GAN} segmentation model misclassifies more pixels as true, whereas \textbf{Styled-Diff} is more conservative in predicting polyp locations. The results for \(r = 1, 2\) are presented in Supplementary Information B.4, alongside a speculative analysis of the effect of cluster size on segmentation performance.}

\begin{table}
  \centering
  \caption{Performance evaluation of the segmentation model for all clusters when small training sets are augmented by synthetic images. Augmentation  strategy: with probability 0.5, replace each real image \hl{with \(r=3\) synthetic images.}}
  \label{table:small_augmentation}
  \begin{tabular}{lccccccccccc}
    \toprule
    \(N\) & Data type & & & \multicolumn{2}{c}{HyperKvasir} & \multicolumn{2}{c}{\hl{CVC-ClinicDB}} & \multicolumn{2}{c}{\hl{PolypGen}} \\
    \cmidrule(r){5-6} \cmidrule(r){7-8} \cmidrule(r){9-10}
     & & & & Dice Score (\(\uparrow\)) & IoU (\(\uparrow\)) & \hl{Dice Score} & \hl{IoU} & \hl{Dice Score} & \hl{IoU}\\
    \midrule
    16 & Real & & & 0.3944 & 0.5263 & 0.3078 & 0.3274 & 0.3395 & 0.4054 \\
    & \hl{10-Styled-Diff} & & & 0.5134 & \textbf{0.6327} & 0.3939 & 0.4722 & 0.3527 & \textbf{0.4601} \\
    & 20-Styled-Diff & & & 0.4529 & 0.6072 & 0.3427 & 0.4971 & 0.3438 & 0.4414 \\
    & \hl{30-Styled-Diff} & & & \textbf{0.5818} & 0.6259 & \textbf{0.4642} & \textbf{0.5160} & 0.3853 & 0.4388 \\
    & \hl{40-Styled-Diff} & & & 0.5467 & 0.6279 & 0.3579 & 0.4531 & 0.3147 & 0.4282 \\
    & GAN & & & 0.4757 & 0.5863 & 0.3852 & 0.4702 & \textbf{0.3933} & 0.4245 \\
    \midrule

    32 & Real & & & 0.4420 & 0.6201 & 0.4160 & 0.5661 & 0.3110 & 0.4497 \\
    & \hl{10-Styled-Diff} & & & 0.6184 & 0.6456 & \textbf{0.5331} & 0.5664 & 0.3401 & 0.4675 \\
    & 20-Styled-Diff & & & 0.7144 & \textbf{0.6812} & 0.4538 & 0.5716 & 0.4301 & \textbf{0.4680} \\
    & \hl{30-Styled-Diff} & & & 0.6099 & 0.6518 & 0.5113 & \textbf{0.5801} & 0.4002 & 0.4610 \\
    & \hl{40-Styled-Diff} & & & 0.5331 & 0.6175 & 0.4403 & 0.5174 & \textbf{0.4453} & 0.4570 \\
    & GAN & & & \textbf{0.7305} & 0.6765 & 0.5311 & 0.5247 & 0.3070 & 0.3979 \\
    \midrule

    64 & Real & & & 0.5799 & 0.6930 & 0.4889 & 0.5729 & \textbf{0.4852} & 0.4768  \\
    & \hl{10-Styled-Diff} & & & 0.6244 & 0.6867 & 0.4175 &\textbf{ 0.6261} & 0.4162 & \textbf{0.5303} \\
    & 20-Styled-Diff & & & \textbf{0.7813} & \textbf{0.7294} & 0.5291 & 0.6249 & 0.3902 & 0.4839 \\
    & \hl{30-Styled-Diff} & & & 0.6197 & 0.7087 & 0.5886 & 0.6166 & 0.3898 & 0.4825 \\
    & \hl{40-Styled-Diff} & & & 0.6968 & 0.7117 & \textbf{0.5933} & 0.5862 & 0.4084 & 0.4690 \\
    & GAN & & & 0.7507 & 0.7098 & 0.5775 & 0.5573 & 0.3637 & 0.4657 \\
    \midrule

    128 & Real & & & 0.8232 & \textbf{0.7982} & 0.6099 & \textbf{0.6449} & 0.5503 & 0.5760  \\
    & \hl{10-Styled-Diff} & & & 0.8025 & 0.7399 & 0.6817 & 0.6377 & 0.4766 & 0.5568 \\
    & 20-Styled-Diff & & & 0.7773 & 0.7494 & 0.6466 & 0.6052 & 0.5513 & \textbf{0.5762} \\
    & \hl{30-Styled-Diff} & & & \textbf{0.8383} & 0.7744 & 0.6820 & 0.6383 & \textbf{0.5868} & 0.5680 \\
    & \hl{40-Styled-Diff} & & & 0.8142 & 0.7655 & \textbf{0.6940} & 0.6422 & 0.5849 & 0.5349 \\
    & GAN & & & 0.8073 & 0.7511 & 0.5964 & 0.5726 & 0.5730 & 0.5364\\
    \bottomrule
  \end{tabular}
\end{table}

\section{Discussion}\label{sec:Discussion}

In this paper, we explored the use of diffusion models and related techniques to generate synthetic data for medical image segmentation.
Our results suggest that data generated by our pipeline may be used to either: replace real data in the training set, or augment real data in small training sets. In the former case, we saw that the synthetic data performs better than the SinGAN-Seg generated data \cite{thambawitaSinGANSegSyntheticTraining2022}, and in the latter case, we saw slight improvements over the Real data and GANs. As opposed to manual gathering and labeling of data, our pipeline is fully automated and thus can be used to generate large amounts of data with minimal human effort. In particular, our pipeline may increase the size and quality of small datasets by generating an arbitrary number of synthetic images from one real image. 

\hl{In contrast to SinGAN-Seg \cite{thambawitaSinGANSegSyntheticTraining2022}, our pipeline is also more computationally efficient. Instead of training a model for each image in the dataset, we performed image clustering and trained a model for each cluster. And indeed, training one diffusion model on the entire dataset still yielded competitive results. Moreover,} in contrast to work by \citet{pishvaRePolypFrameworkGenerating2023,duArSDMColonoscopyImages2023}, our method used a fourth channel to generate the segmentation mask. This allowed us to generate the segmentation mask simultaneously with the image, and thus, alleviate the need for extra models to generate the mask. In our work, we also explored the RePaint technique when the polyps were the subject being inpainted onto the source image, as opposed to the method by \citet{pishvaRePolypFrameworkGenerating2023} which inpainted the background. This generally resulted in a more realistic image. We also used a combination of styling techniques to improve the appearance of the synthetic images which was not explored in either work. 

Our study has several limitations. First, we only used one \hl{training} dataset in our paper. The application of our technique to other, small, low-similarity image segmentation datasets may be done to reinforce the results in this paper. Second, the metrics we used to evaluate the synthetic images are not perfect, especially in the realm of medical images. \hl{For instance, FID is typically used for general image datasets such as ImageNet.} Therefore, caution has to be exercised when interpreting the results. Finally, in this paper, we were constrained by time and computational resources to only explore 3 image samples per real image. Other than the recommendations above, another avenue for future work is the effect of using differentially-private diffusion models \cite{ghalebikesabiDifferentiallyPrivateDiffusion2023}. If such models do not compromise performance, then we may extend our method to a wider range of datasets and enable the use of our pipeline as an image-sharing technique, i.e. instead of original images, we may instead share images generated from them. 

In summary, our pipeline can be used to generate more realistic synthetic data for medical image segmentation. This data can be used to augment small datasets, or replace the real data in the training set. In particular, we see that our pipeline outperforms SinGAN-generated data \cite{thambawitaSinGANSegSyntheticTraining2022} in both cases, and styling sufficiently improves the appearance of the synthetic images to be used in training.

\newpage

\appendixpageoff
\appendixtitleoff
\renewcommand{\appendixtocname}{Supplementary material}
\begin{appendices}

\section*{Supplementary material}

\section{Background}\label{secA1}

\subsection{Background on Diffusion Models}
\label{app:ddpm}
Diffusion models were first formulated by \citet{hoDenoisingDiffusionProbabilistic2020} and later improved by \citet{nicholImprovedDenoisingDiffusion2021}. The diffusion model consists of two primary components: the forward diffusion process, denoted \(q\), and the reverse denoising process, denoted \(p\). Starting at a clean image \(x_0\), we gradually add Gaussian noise to it for \(T\) timesteps. In particular, \[
  q(x_T|x_0) = \prod_{t = 1}^Tq(x_t|x_{t-1})
\]
where \[
  q(x_t|x_{t-1}) = \mathcal{N}(x_t; \sqrt{1-\beta_t}x_{t-1}, \beta_t(1-\beta_t)I),
\]
and where \(\beta_1, \ldots, \beta_T\) is the variance schedule (e.g. the one proposed by \citet{nicholImprovedDenoisingDiffusion2021}), and \(x_T\) is the noisy image. We can however reparameterize the distribution by letting \(\alpha_t := 1 - \beta_t\) and \(\bar \alpha_t := \prod_{r = 1}^t\alpha_r\). Then, we may sample any \(t \in \{1, \ldots, T\}\) directly by 
\begin{equation}
\label{eq:diffusion_sample}
  x_t = \sqrt{\bar \alpha_t}x_0 + \sqrt{1 - \bar \alpha_t}\epsilon_t, \quad \epsilon_t \sim \mathcal{N}(0, I).
\end{equation}

The reverse denoising process is the process of denoising the noisy image \(x_T\) to recover the clean image \(x_0\). This process may only be approximated, and is the component of the model which is learned. In particular, we learn to maximize the variational lower bound by gradual transitions \(p_\theta(x_{t-1} | x_t)\) determined by the model parameters \(\theta\). Starting at \(p(x_T) = \mathcal{N}(x_T; 0, I) \), we obtain \[
  p_\theta(x_{0:T}) = p(x_T)\prod_{t=1}^Tp_\theta(x_{t-1}|x_t),
\]
where \[
  p_\theta(x_{t-1}|x_t) = \mathcal{N}(x_{t-1}; \mu_\theta(x_t, t), \Sigma_\theta(x_t, t)).
\]
The parameters above are what we would like to learn. \citet{nicholImprovedDenoisingDiffusion2021} described a method to learn \(\Sigma_\theta(x_t, t)\), which we employ in our training. However, most important is \[
  \mu_\theta(x_t, t) := \frac{1}{\sqrt{\bar \alpha_t}}\left(x_t - \frac{1-\alpha_t}{\sqrt{1 - \bar \alpha_t}}\epsilon_\theta(x_t, t)\right).
\]
\citet{hoDenoisingDiffusionProbabilistic2020} found that it was best to optimize the "noise" term, \(\epsilon_\theta(x_t, t)\). In particular, the objective is \[
  \min_\theta \mathbb{E}_{t, x_0, \epsilon}\left[\|\epsilon - \epsilon_\theta(x_t, t)\|^2\right].
\] We may then predict \(x_{t-1}\) by 
\begin{equation}
\label{eq:denoised_sample}
  x_{t - 1} = \mu_\theta(x_t, t) + \sigma_t\epsilon,
\end{equation}
and henceforth obtain the final prediction \(\hat x_0\).

\subsection{Using Diffusion Models for Image Inpainting}
\label{app:inpainting}
The technique of inpainting with diffusion models were first introduced by \citet{lugmayrRePaintInpaintingUsing2022}. The approach involves a mask \(m\) with pixel values either 0 or 1, where 0 indicates the unknown regions, i.e. regions we would like to paint in, and 1 indicates known regions, i.e. regions we would like to maintain. A simple modification of the reverse denoising process is needed to achieve this goal. Suppose we wish to obtain \(x_{t- 1}\) from \(x_t\). We only want the denoising process to take place on the unknown region of the mask, hence we use Equation \ref{eq:diffusion_sample} to obtain the known regions: \(x_{t-1}^{known} = m \odot x_{t-1}\), and then use Equation \ref{eq:denoised_sample} to obtain the unknown regions: \(x_{t-1}^{unknown} = (1 - m) \odot x_{t-1}\). We then combine the two:
\begin{equation}
\label{eq:inpainting}
x_{t-1} = x_{t-1}^{known} + x_{t-1}^{unknown}.
\end{equation}

\subsection{Using Diffusion Models for Image Styling}
\label{app:styling}
In our pipeline, the technique of using diffusion models for styling is based on the work of \citet{kwonDiffusionbasedImageTranslation2023,yangZeroShotContrastiveLoss2023a}. The method works by having a designated target image, which we use to guide the translation of the source image. In our case, the source image will be the image generated via inpainting by our diffusion model, and the target image is the original image which was inpainted. The objective of image translation is to preserve the structural content, while varying the semantic content to better match the target. As such, we have two general sets of losses: content and style loss. These conditional losses are applied at each reverse timestep \(t\). In particular, if we let \(\ell_{total}\) be the style loss, then like in Equation \ref{eq:denoised_sample}, we obtain \begin{align*}
  x_{t-1}' &= \mu_\theta(x_t, t) + \sigma_t\epsilon, \quad \epsilon \sim \mathcal{N}(0, I)\\
  x_{t-1} &= x_{t-1}'  - \nabla_{x_t}\ell_{total}({\hat{x}}_0(x_t)),
\end{align*}
where \(\hat x_0(x_t)\) is the predicted clean image from the noisy sample at timestep \(t\) obtained using Tweedie's formula \cite{kimNoise2ScoreTweedieApproach2021}. For our convenience, we write \(x := \hat x_0(x_t)\).

We first focus our attention to the content losses. In this case, we make use of the losses in \cite{yangZeroShotContrastiveLoss2023a}: \(\ell_{ZeCon}, \ell_{VGG}\) The ZeCon loss takes in \(x, x_0\) -- note that \(x_0\) is the target image -- and forwards them to the UNet, i.e. the estimator of \(\epsilon_\theta(x_t, t)\). The UNet encoder is used to extract feature maps at some layer \(l\) of both \(x\) and \(x_t\); we denote their respective feature maps \(\hat z_l\) and \(z_l\). We then randomly select a spatial location on the feature \(z_l\) from the set of all locations \(s \in \{1, \ldots, S_l\} =: S\) on which we compute the cross-entropy loss. The Zecon loss is then 
\begin{equation}
\label{eq:zecon}
\ell_{ZeCon}(x, x_0) := \mathbb{E}_{x_0} \left[ \sum_l \sum_s \mathcal{L}_{CE}(\hat z_l^s, z_l^s, z_l^{S \setminus s}) \right]
\end{equation}
The VGG loss is defined simply as the MSE between VGG feature maps of \(x, x_0\).

Next, we deal with the style losses which we take adapt from \cite{kwonDiffusionbasedImageTranslation2023}: \(\ell_{sty}, \ell_{L2}, \ell_{sem}, \ell_{rng}\). A primary component of our losses is the DINO ViT \cite{caronEmergingPropertiesSelfSupervised2021} which was shown by \citet{tumanyanSplicingViTFeatures2022} to delineate between structural content and semantic content. In particular, the [CLS] token of the last layer was shown to contain semantic information; we will follow \citet{kwonDiffusionbasedImageTranslation2023} and denote it \(e^L_{[CLS]}(\cdot)\).

The primary semantic style loss we use is \(\ell_{sty}\):
\begin{equation}
\label{eq:sty}
\ell_{sty}(x, x_0) := \|e^L_{[CLS]}(x_0) - e^L_{[CLS]}(x)\|_2.
\end{equation}
In order to accelerate the generation process, we wish to maximize the semantic distance between the estimated clean images of adjacent timesteps. Therefore, we define \(\ell_{sem}\) as follows:
\begin{equation}
\label{eq:sem}
\ell_{sem}(x_t, x_{t-1}) := -\|e^L_{[CLS]}(\hat x_0(x_{t-1})) - e^L_{[CLS]}(\hat x_0(x_t))\|_2.
\end{equation}
The losses \(\ell_{L2}, \ell_{rng}\) are defined more simply: the former is just the MSE of the pixel difference between \(x, x_0\), and the latter is a regularization loss to prevent irregular steps in the reverse process. We thus have:
\begin{align*}
  \mathcal{L}_{content} &= \lambda_{ZeCon}\ell_{ZeCon}(x, x_0) + \lambda_{VGG}\ell_{VGG}(x, x_0) \\
  \mathcal{L}_{style} &= \lambda_{sty}\ell_{sty}(x, x_{src}) + \lambda_{L2}\ell_{L2}(x, x_{src}) + \lambda_{sem}\ell_{sem}(x_t; x_{t-1}) + \lambda_{rng}\ell_{rng} \\
  \mathcal{L}_{total} &= \mathcal{L}_{content} + \mathcal{L}_{style}
\end{align*}
where \(x_0\) is the inpainted image and \(x_{src}\) is the real image which was inpainted. 

\section{Further Methods and Results} \label{secA2}
\subsection{Expert Review}
\label{app:expert_review}
Three assessments were made: image quality, segmentation quality, and whether the image was real or synthetic. The image and segmentation quality assessments had 3 options: Good, Moderate, and Poor. Using this data, accuracy, recall, and precision were calculated to measure the agreement in the classification of real vs. fake images.

The results of the expert review are shown in Table \ref{table:expert_review_main}. Combining the Diff and Styled-Diff sets as the "Fake" set of images, we obtain an accuracy, precision, and recall of 60\%, 40\%, and 40\% respectively. Moreover, this review confirms that the generations are indeed good quality data points, although some may identified as synthetic. It is also interesting to see how the real images were rated since it was the category with the highest number of fake images identified.

\begin{table}[h]
  \centering
  \caption{Survey results from an expert rater tasked with assessing image quality, segmentation quality, and whether the image was real or synthetic.}
  \label{table:expert_review_main}
  \begin{tabular}{ccccccccccc}
    \toprule
     & & & \multicolumn{3}{c}{Image Quality} & \multicolumn{3}{c}{Segmentation Quality} & \multicolumn{2}{c}{Real or Fake} \\
    \cmidrule(r){4-6} \cmidrule(r){7-9} \cmidrule(r){10-11}
    Data type & & & Good & Moderate & Poor & Good & Moderate & Poor & Real & Fake \\
    \midrule
    Real & & & 10 & 0 & 0 & 8 & 2 & 0 & 4 & 6 \\
    Diff & & & 10 & 0 & 0 & 10 & 0 & 0 & 3 & 7 \\
    Styled-Diff & & & 10 & 0 & 0 & 9 & 1 & 0 & 3 & 7 \\
    \bottomrule
  \end{tabular}
\end{table}

\subsection{Intrinsic Evaluation}
\label{app:intrinsic}
\hl{The intrinsic evaluation of the synthetic images was done using the FID \cite{heuselGANsTrainedTwo2017} and MS-SSIM \cite{wangMultiscaleStructuralSimilarity2003} score. We computed the FID and MS-SSIM score between the generated images and the real data.
}

\hl{The FID \cite{heuselGANsTrainedTwo2017} and MS-SSIM \cite{wangMultiscaleStructuralSimilarity2003} scores measure the distance between the output images and the dataset, as well as the variance of the output images. \citet{parmarAliasedResizingSurprising2022} showed that differences in image compression and resizing, among other factors, may induce significant variation in the FID scores. Hence, we used the method introduced in \cite{parmarAliasedResizingSurprising2022} to compute the FID. In the main paper we compared the scores between the real and various synthetically generated images using cluster size 20. For completeness, we also computed the FID and MS-SSIM scores between the real and synthetic images using cluster sizes 10, 30, and 40. The results are shown in Table \ref{table:fidnmsssim}.}

\begin{table}
  \centering
  \caption{Fr\'echet Inception Distance (FID) and Multi-Scale Structural Similarity (MS-SSIM) comparison between real and synthetic images. Images are shuffled, and each data type (including \textit{Real}) is compared with \textit{Real}. The computations were performed thrice.}
  \label{table:fidnmsssim}
  \begin{tabular}{cccccccccc}
  \toprule
  Data type & & &\multicolumn{3}{c}{FID} & \multicolumn{3}{c}{MS-SSIM} \\
  \cmidrule(r){4-6}
  \cmidrule(r){7-9}
   & & & Set 1 & Set 2 & Set 3 & Set 1 & Set 2 & Set 3 \\
  \midrule
  10-Diff	& & & 35.86 &	35.82 &	35.90 &	0.2018	& 0.2071 &	0.2078 \\
  20-Diff & & & 36.01 & 37.90 & 37.78  & 0.2081 & 0.2126 & 0.2096 \\
  30-Diff & & &	42.90 &	43.36 &	43.12	& 0.2117 &	0.2259 &	0.2204\\
  40-Diff	& & & 44.93 &	44.70	& 45.40 &	0.2050 &	0.2004 & 0.1984 \\
  \midrule
  10-Styled-Diff & & & 65.71 &	65.09 &	65.93 &	0.2158 &	0.2098 &	0.2206 \\
  20-Styled-Diff	& & &	66.17 & 65.48 & 66.32 & 0.2304 & 0.2300 & 0.2433	\\			
  30-Styled-Diff	& & & 96.53 &	98.04 &	95.11 &	0.2704 &	0.2675 &	0.2686 \\
  40-Styled-Diff	& & & 91.94 &	90.56 &	92.24 &	0.2743 &	0.2670 &	0.2609 \\
  \bottomrule
  \end{tabular}
\end{table}

\hl{It is interesting to note the sudden increase in FID and MS-SSIM scores from \(K = 20\) to \(K = 30, \, 40\). A plausible explanation is that for higher cluster counts, the models have a more limited set of polyps to inpaint onto the source images. Recalling that the background and the polyp come from different clusters (Section 2.4), the styling process, which is guided by the source image, is significantly more transformative. Hence, the generated images are more varied and less similar to the real images. In contrast, for smaller cluster counts, the greater cluster size means that the diffusion models have a larger set of polyps to inpaint onto the source images. Thus, the inpainting model can select a polyp that better fits the source image, which is often similar to the real image. As we shall see, this deviation from the real image is not necessarily a downside, as the synthetic images are still useful for training segmentation models.}

\subsection{Segmentation Model Training Details}
\label{app:segt_model}
The model used in \cite{thambawitaDivergentNetsMedicalImage2021} employs a UNet++ backbone. Specifically, the model was trained only using the generated data and tested using real data. We used the \verb|se_resnext50_32x4d| network as the UNet++ encoder and \verb|softmax2d| as the last layer activation function. PyTorch was used as the devlopment framework, and the data stream was handled by PYRA along with the Albumentations augmentation library. The standard training augmentations used were: \verb|ShiftScaleRotate|, \verb|HorizontalFlip|, and one of \verb|CLAHE, RandomBrightness, RandomGamma|. The model weights were zero-initialized and they were trained for 150 epochs, with a learning rate of 0.0001 for the first 50 epochs, which was reduced to 0.00001 for the remaining epochs.

\begin{table}
  \centering
  \caption{Performance evaluation of segmentation models entirely trained on synthetic data. The test datasets are real images not used in the generation.}
  \label{table:segmentation}
  \begin{tabular}{ccccccccc}
    \toprule
     & & & \multicolumn{2}{c}{HyperKvasir} & \multicolumn{2}{c}{{CVC-ClinicDB}} & \multicolumn{2}{c}{{PolypGen}} \\
    \cmidrule(r){4-5} \cmidrule(r){6-7} \cmidrule(r){8-9}
    Data type & & & Dice Score & IoU & {Dice Score} & {IoU} & {Dice Score} & {IoU}\\
    \midrule  
    Real & & & 0.8680 & 0.8085 & 0.7341 & 0.6604 & 0.6185 & 0.6284\\
    \midrule

    10-Diff-1 & & & \textbf{0.7036} &\textbf{ 0.6393} & \textbf{0.5919} & 0.5268 & 0.4322 & 0.4875\\
    20-Diff-1 & & & 0.6566 & 0.5948 & 0.5845 & \textbf{0.5279} & \textbf{0.4351} & \textbf{0.4923} \\
    30-Diff-1 & & & 0.6310 & 0.5705 & 0.4217 & 0.3780 & 0.3418 & 0.3861 \\
    40-Diff-1 & & & 0.4771 & 0.4233 & 0.3751 & 0.3391 & 0.3608 & 0.4187 \\
    \midrule

    10-Diff-2 & & & \textbf{0.7453 }& \textbf{0.6703} & \textbf{0.6333} & 0.5684 & 0.4280 & \textbf{0.5238} \\
    20-Diff-2 & & & 0.6626 & 0.5975 & 0.6315 & \textbf{0.5827} & \textbf{0.4934} & 0.5203 \\
    30-Diff-2 & & & 0.6059 & 0.5418 & 0.4289 & 0.3810 & 0.3019 & 0.3443 \\
    40-Diff-2 & & & 0.5562 & 0.4967 & 0.4979 & 0.4419 & 0.4022 & 0.4246 \\
    \midrule
    
    10-Diff-3 & & & \textbf{0.7125} & \textbf{0.6360} & 0.6206 & 0.5562 & \textbf{0.5035} & \textbf{0.5286} \\
    20-Diff-3 & & & 0.6516 & 0.5883 & \textbf{0.6335} & \textbf{0.5776} & 0.4694 & 0.4703\\
    30-Diff-3 & & & 0.5892 & 0.5212 & 0.4332 & 0.3901 & 0.4274 & 0.4271 \\
    40-Diff-3 & & & 0.6120 & 0.5512 & 0.4112 & 0.3884 & 0.3694 & 0.4086 \\
    \bottomrule
  \end{tabular}
\end{table}

\begin{table}
  \centering
  \caption{Performance evaluation of segmentation models entirely trained on styled synthetic data. {The test datasets are real images not used in the generation.}}
  \label{table:segmentation_styled}
  \begin{tabular}{ccccccccc}
    \toprule
     & & & \multicolumn{2}{c}{HyperKvasir} & \multicolumn{2}{c}{{CVC-ClinicDB}} & \multicolumn{2}{c}{{PolypGen}} \\
    \cmidrule(r){4-5} \cmidrule(r){6-7} \cmidrule(r){8-9}
    Data type & & & Dice Score & IoU & {Dice Score} & {IoU} & {Dice Score} & {IoU}\\
    \midrule  
    Real & & & 0.8680 & 0.8085 & 0.7341 & 0.6604 & 0.6185 & 0.6284\\
    \midrule

    10-Styled-Diff-1 & & & \textbf{0.7596} & \textbf{0.6991} & \textbf{0.6692} & \textbf{0.6231} & 0.4996 & 0.5544\\
    20-Styled-Diff-1 & & & 0.7444 & 0.6840 & 0.6363 & 0.6027 & \textbf{0.5669} & \textbf{0.5871}\\
    30-Styled-Diff-1 & & & 0.7412 & 0.6821 & 0.5771 & 0.5764 & 0.5131 & 0.5382\\
    40-Styled-Diff-1 & & & 0.6676 & 0.6341 & 0.5758 & 0.5347 & 0.4448 & 0.4601\\
    \midrule

    10-Styled-Diff-2 & & & 0.7212 & 0.6700 & 0.5972 & 0.5403 & 0.5125 & 0.5742\\
    20-Styled-Diff-2 & & & \textbf{0.7728} & \textbf{0.7027} & \textbf{0.6414} & \textbf{0.5890} & 0.5444 & \textbf{0.5752}\\
    30-Styled-Diff-2 & & & 0.6989 & 0.6442 & 0.6338 & 0.5764 & \textbf{0.5668} & 0.5708\\
    40-Styled-Diff-2 & & &  0.7479 & 0.6924 & 0.5834 & 0.5194 & 0.4731 & 0.4782\\
    \midrule

    10-Styled-Diff-3 & & & 0.7894 & 0.7277 & \textbf{0.7396} & \textbf{0.6643 }& 0.5413 & 0.5542\\
    20-Styled-Diff-3 & & & \textbf{0.8042} & \textbf{0.7396} & 0.6749 & 0.6124 & \textbf{0.5753} & \textbf{0.5728} \\
    30-Styled-Diff-3 & & & 0.7540 & 0.6936 & 0.6679 & 0.6106 & 0.5240 & 0.5217\\
    40-Styled-Diff-3 & & & 0.6320 & 0.5663 & 0.6020 & 0.5408 & 0.4568 & 0.4592\\
    \bottomrule
  \end{tabular}
\end{table}

\subsection{Segmentation Experiments and Transfer Learning}
\label{app:transfer}
\subsubsection{Full Synthetic Training}
\hl{
The segmentation model was trained on the synthetic images generated by our pipeline. The model was then tested on the real images from HyperKvasir, CVC-ClinicDB, and PolypGen. The results of the segmentation model are shown in Table \ref{table:segmentation} and Table \ref{table:segmentation_styled}. It is clear to see that the model trained on fewer clusters performed better than the model trained on more clusters, at times differing by more than \(0.2\) in both metrics. This continues to support the hypothesis that the model trained on fewer clusters has a more diverse set of polyps to inpaint onto the source images, which results in a more realistic set of synthetic images. Indeed, this improved performance is seen on all test sets.}

\subsubsection{Augmentation of Small Datasets}
\hl{In this experiment, we augment small datasets with our generated data. The results are given in Tables \ref{table:small_augmentation1}, \ref{table:small_augmentation2}, \ref{table:small_augmentation3}, each differing by the number of synthetic images added per augmentation. With the other cluster counts considered the \textbf{Styled-Diff} models performed better than the \textbf{GAN} models, with only one or two exceptions per Table.}

\hl{Although the pattern is not as clear, it is interesting to note the performance of the larger cluster counts in this experiment. A plausible explanation is that the greater variation in the synthetic images, although unrealistic, is beneficial for enhancing segmentation models with small training sets. This is especially plausible in this experiment because we have a reliable base of real images along with the synthetic images. This is further supported by the fact that for \(N = 128\), \textbf{30/40-Styled-Diff} models often outperformed \textbf{10/20-Styled-Diff} models, or even the \textbf{Real} model. In contrast, when only synthetic images were used, the \textbf{10/20-Styled-Diff} models performed better than the \textbf{30/40-Styled-Diff} models, as shown in Table \ref{table:segmentation_styled}.}

\hl{It is therefore possible to say that the choice of cluster count is dependent on the use case. If we are using the synthetic images to fully train a machine learning model, or image-sharing, the more realistic images generated by the smaller cluster counts would be more beneficial. However, if we are using the synthetic images to augment a dataset which lacks image diversity, the larger cluster counts, with its greater variability, would be more beneficial.}

\begin{table}
  \centering
  \caption{Performance evaluation of the segmentation model when small training sets are augmented by synthetic images. Augmentation  strategy: with probability 0.5, replace each real image with \(r=1\) synthetic images.}
  \label{table:small_augmentation1}
  \begin{tabular}{lccccccccccc}
    \toprule
    \(N\) & Data type & & & \multicolumn{2}{c}{HyperKvasir} & \multicolumn{2}{c}{CVC-ClinicDB} & \multicolumn{2}{c}{PolypGen} \\
    \cmidrule(r){5-6} \cmidrule(r){7-8} \cmidrule(r){9-10}
     & & & & Dice Score & IoU & Dice Score & IoU & Dice Score & IoU\\
    \midrule
    16 & Real & & & 0.3944 & 0.5263 & 0.3078 & 0.3274 & 0.3395 & 0.4054\\
    & 10-Styled-Diff & & & 0.4948 & 0.6160 & 0.3891 & 0.4367 & \textbf{0.3747} & 0.4022 \\
    & 20-Styled-Diff & & & 0.5162 & 0.5930 & 0.3587 & \textbf{0.4855} & 0.3694 & \textbf{0.4517} \\
    & 30-Styled-Diff & & & \textbf{0.5662} & \textbf{0.6462} & \textbf{0.4061} & 0.4642 & 0.3272 & 0.4299 \\
    & 40-Styled-Diff & & & 0.4657 & 0.5804 & 0.3370 & 0.4303 & 0.2625 & 0.4062 \\
    & GAN & & & 0.4835 & 0.5658 & 0.3778 & 0.4625 & 0.3727 & 0.4175  \\

    \midrule
    32 & Real & & & 0.4420 & 0.6201 & 0.4160 & \textbf{0.5661} & 0.3110 & 0.4497 \\
    & 10-Styled-Diff & & & 0.6370 & 0.5907 & 0.5499 & 0.4869 & 0.4447 & 0.3370 \\
    & 20-Styled-Diff & & & 0.5634 & \textbf{0.6494} & 0.3522 & 0.5604 & 0.3634 & \textbf{0.4610}\\
    & 30-Styled-Diff & & & 0.6569 & 0.5918 & \textbf{0.5782} & 0.3920 & \textbf{0.4779} & 0.4229 \\
    & 40-Styled-Diff & & & \textbf{0.6662} & 0.5656 & 0.5475 & 0.3848 & 0.4559 & 0.4037 \\
    & GAN & & &  0.5864 & 0.6326 & 0.3899 & 0.5143 & 0.3153 & 0.4301 \\
    \midrule

    64 & Real & & & 0.5799 & 0.6930 & 0.4889 & 0.5729 & \textbf{0.4852} & 0.4768 \\
    & 10-Styled-Diff & & & 0.6543 & 0.7033 & 0.5067 & 0.6012 & 0.4037 & 0.4790 \\
    & 20-Styled-Diff & & & 0.7155 & \textbf{0.7262} & 0.3931 & 0.5890 & 0.4285 & \textbf{0.4950} \\
    & 30-Styled-Diff & & & 0.6305 & 0.6772 & 0.4137 & \textbf{0.6203} & 0.3439 & 0.4826 \\
    & 40-Styled-Diff & & & 0.5590 & 0.6974 & \textbf{0.5096} & 0.6102 & 0.3126 & 0.4644 \\
    & GAN & & & \textbf{0.7580} &  0.7184 & 0.4244 & 0.5534 & 0.3119 & 0.4314 \\
    \midrule

    128 & Real & & & 0.8232 & \textbf{0.7982} & 0.6099 & 0.6449 & \textbf{0.5503} & \textbf{0.5760}  \\
    & 10-Styled-Diff & & & 0.7659 & 0.7686 & 0.5373 & \textbf{0.6499} & 0.5336 & 0.5479 \\
    & 20-Styled-Diff & & & 0.8122 & 0.7511 & 0.5148 & 0.6461 & 0.4769 & 0.5584 \\
    & 30-Styled-Diff & & & 0.7688 & 0.7476 & \textbf{0.6480} & 0.6425 & 0.5100 & 0.5436 \\
    & 40-Styled-Diff & & & \textbf{0.8246} & 0.7576 & 0.5679 & 0.6170 & 0.4367 & 0.5297 \\
    & GAN & & & 0.8230 & 0.7536 & 0.4755 & 0.5906 & 0.4576 & 0.5311 \\
    \bottomrule
  \end{tabular}
\end{table}

\begin{table}
  \centering
  \caption{Performance evaluation of the segmentation model when small training sets are augmented by synthetic images. Augmentation  strategy: with probability 0.5, replace each real image with \(r=2\) synthetic images.}
  \label{table:small_augmentation2}
  \begin{tabular}{lccccccccccc}
    \toprule
    \(N\) & Data type & & & \multicolumn{2}{c}{HyperKvasir} & \multicolumn{2}{c}{CVC-ClinicDB} & \multicolumn{2}{c}{PolypGen} \\
    \cmidrule(r){5-6} \cmidrule(r){7-8} \cmidrule(r){9-10}
     & & & & Dice Score & IoU & Dice Score & IoU & Dice Score & IoU\\
    \midrule

    16 & Real & & & 0.3944 & 0.5263 & 0.3078 & 0.3274 & 0.3395 & 0.4054\\
    & 10-Styled-Diff & & & 0.5515 & 0.6276 & 0.3758 & 0.4288 & 0.3605 & 0.4576 \\
    & 20-Styled-Diff & & & 0.5114 & 0.5968 & 0.3277 & 0.4620 & 0.3610 & \textbf{0.4675} \\
    & 30-Styled-Diff & & & 0.4879 & 0.6112 & 0.3830 & \textbf{0.4855} & \textbf{0.3891} & 0.4310 \\
    & 40-Styled-Diff & & & \textbf{0.5554} & \textbf{0.6400} & 0.3209 & 0.4496 & 0.3212 & 0.4147 \\
    & GAN & & & 0.4997 & 0.5762 & \textbf{0.4002} & 0.4735 & 0.3146 & 0.4020 \\

    \midrule
    32 & Real & & & 0.4420 & 0.6201 & 0.4160 & 0.5661 & 0.3110 & 0.4497 \\
    & 10-Styled-Diff & & & 0.5605 & 0.6496 & 0.3986 & \textbf{0.5916} & \textbf{0.4052} & \textbf{0.4697} \\
    & 20-Styled-Diff & & & \textbf{0.6605} & \textbf{0.6690} & 0.3498 & 0.5801 & 0.2572 & 0.4311 \\
    & 30-Styled-Diff & & & 0.5510 & 0.6668 & \textbf{0.5414} & 0.5817 & 0.2743 & 0.4426 \\
    & 40-Styled-Diff & & & 0.5466 & 0.6534 & 0.3740 & 0.5561 & 0.2871 & 0.4337 \\
    & GAN & & & 0.6587 & 0.6609 & 0.4430 & 0.5791 & 0.3670 & 0.4269 \\
    \midrule

    64 & Real & & & 0.5799 & 0.6930 & 0.4889 & 0.5729 & \textbf{0.4852} & 0.4768 \\
    & 10-Styled-Diff & & & 0.5989 & 0.6988 & 0.4503 & \textbf{0.6273} & 0.4351 & 0.5138 \\
    & 20-Styled-Diff & & & 0.7149 & 0.7133 & 0.5191 & 0.5804 & 0.4146 & 0.4966\\
    & 30-Styled-Diff & & &  0.6783 & 0.7037 & 0.4316 & 0.6050 & 0.4345 & \textbf{0.5167} \\
    & 40-Styled-Diff & & & \textbf{0.7285} & 0.7010 & \textbf{0.5389} & 0.6112 & 0.3843 & 0.4923 \\
    & GAN & & &  0.7024 & \textbf{0.7142} & 0.4129 & 0.5874 & 0.3743 & 0.4456 \\
    \midrule
    128 & Real & & & \textbf{0.8232} & \textbf{0.7982} & 0.6099 & 0.6449 & \textbf{0.5503} & \textbf\textbf{0.5760}  \\
    & 10-Styled-Diff & & & 0.7960 & 0.7313 & \textbf{0.6802} & 0.6444 & 0.4476 & 0.5458 \\
    & 20-Styled-Diff & & & 0.7719 & 0.7525 & 0.6308 & 0.6407 & 0.4259 & 0.5748 \\
    & 30-Styled-Diff & & & 0.8217 & 0.7621 & 0.5455 & \textbf{0.6720} & 0.4766 & 0.5024 \\
    & 40-Styled-Diff & & & 0.7913 & 0.7598 & 0.6719 & 0.6264 & 0.4979 & 0.5387 \\
    & GAN & & &  0.8100 & 0.7513 & 0.6510 & 0.5844 & 0.5161 & 0.5262\\
    \bottomrule
  \end{tabular}
\end{table}

\begin{table}
  \centering
  \caption{Performance evaluation of the segmentation model when small training sets are augmented by synthetic images. Augmentation  strategy: with probability 0.5, replace each real image {with \(r=3\) synthetic images.}}
  \label{table:small_augmentation3}
  \begin{tabular}{lccccccccccc}
    \toprule
    \(N\) & Data type & & & \multicolumn{2}{c}{HyperKvasir} & \multicolumn{2}{c}{{CVC-ClinicDB}} & \multicolumn{2}{c}{{PolypGen}} \\
    \cmidrule(r){5-6} \cmidrule(r){7-8} \cmidrule(r){9-10}
     & & & & Dice Score & IoU & {Dice Score} & {IoU} & {Dice Score} & {IoU}\\
    \midrule
    16 & Real & & & 0.3944 & 0.5263 & 0.3078 & 0.3274 & 0.3395 & 0.4054 \\
    & 10-Styled-Diff & & & 0.5134 & \textbf{0.6327} & 0.3939 & 0.4722 & 0.3527 & \textbf{0.4601} \\
    & 20-Styled-Diff & & & 0.4529 & 0.6072 & 0.3427 & 0.4971 & 0.3438 & 0.4414 \\
    & 30-Styled-Diff & & & \textbf{0.5818} & 0.6259 & \textbf{0.4642} & \textbf{0.5160} & 0.3853 & 0.4388 \\
    & 40-Styled-Diff & & & 0.5467 & 0.6279 & 0.3579 & 0.4531 & 0.3147 & 0.4282 \\
    & GAN & & & 0.4757 & 0.5863 & 0.3852 & 0.4702 & \textbf{0.3933} & 0.4245 \\
    \midrule

    32 & Real & & & 0.4420 & 0.6201 & 0.4160 & 0.5661 & 0.3110 & 0.4497 \\
    & 10-Styled-Diff & & & 0.6184 & 0.6456 & \textbf{0.5331} & 0.5664 & 0.3401 & 0.4675 \\
    & 20-Styled-Diff & & & 0.7144 & \textbf{0.6812} & 0.4538 & 0.5716 & 0.4301 & \textbf{0.4680} \\
    & 30-Styled-Diff & & & 0.6099 & 0.6518 & 0.5113 & \textbf{0.5801} & 0.4002 & 0.4610 \\
    & 40-Styled-Diff & & & 0.5331 & 0.6175 & 0.4403 & 0.5174 & \textbf{0.4453} & 0.4570 \\
    & GAN & & & \textbf{0.7305} & 0.6765 & 0.5311 & 0.5247 & 0.3070 & 0.3979 \\
    \midrule

    64 & Real & & & 0.5799 & 0.6930 & 0.4889 & 0.5729 & \textbf{0.4852} & 0.4768  \\
    & 10-Styled-Diff & & & 0.6244 & 0.6867 & 0.4175 &\textbf{ 0.6261} & 0.4162 & \textbf{0.5303} \\
    & 20-Styled-Diff & & & \textbf{0.7813} & \textbf{0.7294} & 0.5291 & 0.6249 & 0.3902 & 0.4839 \\
    & 30-Styled-Diff & & & 0.6197 & 0.7087 & 0.5886 & 0.6166 & 0.3898 & 0.4825 \\
    & 40-Styled-Diff & & & 0.6968 & 0.7117 & \textbf{0.5933} & 0.5862 & 0.4084 & 0.4690 \\
    & GAN & & & 0.7507 & 0.7098 & 0.5775 & 0.5573 & 0.3637 & 0.4657 \\
    \midrule

    128 & Real & & & 0.8232 & \textbf{0.7982} & 0.6099 & \textbf{0.6449} & 0.5503 & 0.5760  \\
    & 10-Styled-Diff & & & 0.8025 & 0.7399 & 0.6817 & 0.6377 & 0.4766 & 0.5568 \\
    & 20-Styled-Diff & & & 0.7773 & 0.7494 & 0.6466 & 0.6052 & 0.5513 & \textbf{0.5762} \\
    & 30-Styled-Diff & & & \textbf{0.8383} & 0.7744 & 0.6820 & 0.6383 & \textbf{0.5868} & 0.5680 \\
    & 40-Styled-Diff & & & 0.8142 & 0.7655 & \textbf{0.6940} & 0.6422 & 0.5849 & 0.5349 \\
    & GAN & & & 0.8073 & 0.7511 & 0.5964 & 0.5726 & 0.5730 & 0.5364\\
    \bottomrule
  \end{tabular}
\end{table}

\subsection{Hyperparameters}
\label{app:hyperparameters}

\hl{For our implementation of the diffusion model, the images were resized to be \(256\times 256\) and the diffusion model had 4 input channels (+1 to accomodate the mask). A batch size of \(128\) was used alongside an AdamW optimizer with a learning rate of \(5 \times 10^{-5}\) throughout training. During inference, our sampling is set to take \(1000\) diffusion steps, however, we used timestep respacing so that overall we sampled over \(T = 200\) timesteps, with us skipping the first \(80\) (respaced) timesteps.}  Moreover, when inpainting the image, we set the jump length \(j = 10\) with \(r = 5\) times resampling; these parameters are described further in \citet{lugmayrRePaintInpaintingUsing2022}. For styling, we used \(\lambda_{ZeCon} = 500, \lambda_{VGG} = 100, \lambda_{MSE} = 5000, \lambda_{sty} = 10000, \lambda_{L2} = 10000, \lambda_{sem} = 40000, \lambda_{rng} = 200 \). For the ViT model used to compute the style and semantic losses, we used the DINO-ViT model \cite{tumanyanSplicingViTFeatures2022} which used the same settings as in \citet{kwonDiffusionbasedImageTranslation2023}.




\end{appendices}


\bibliography{rop-bib}

\bigskip


\bibliography{rop-bib}

\end{document}